\documentstyle[12pt,epsfig,rotating]{article}
\newcommand{\average}[1]{\mbox{$ \langle #1 \rangle $}}

\newcommand{\GeVx}{\rm GeV}

\newcommand{\gsp}{\mbox{$\gamma^{*}p$}}

\newcommand{\gsim}{\raisebox{-0.5mm}{$\stackrel{>}{\scriptstyle{\sim}}$}}

\begin{document}
\begin{titlepage}
\begin{flushleft}
%===> Change report numbers and date
%
%
{\tt hep-ep/9508013}\\

\vspace{1cm}

{\tt DESY 95-156    \hfill    ISSN 0418-9833} \\
{\tt August 1995}\\
\end{flushleft}
\vspace*{4.cm}
\begin{center}
\begin{Large}
\boldmath
\bf{Comparison of Deep Inelastic Scattering with Photoproduction Interactions
at HERA\\}
\unboldmath

\vspace*{2.cm}
H1 Collaboration \\
\end{Large}

\vspace*{1cm}

\end{center}

\vspace*{1cm}

\begin{abstract}
\noindent
  Photon-proton ($\gamma p$) interactions with~$Q^2<10^{-2}$~GeV$^2$ and
deep-inelastic scattering ($\gamma^* p$) interactions with photon virtualities
$Q^2>5$~GeV$^2$ are studied at the high energy electron-proton collider HERA.
The transverse energy flow and relative rates of large rapidity gap events are
compared in the two event samples. The observed similarity between $\gamma p$
and $\gamma^* p$ interactions can be understood in a picture where the photon
develops as a  hadronic object. The transverse energy density measured in the
central region of the collision, at $\eta^*=0$ in the $\gamma^* p$ centre of
mass
frame, is compared with  data  from hadron-hadron interactions as function of
the
CMS energy of the collision.
\vspace{1cm}

\end{abstract}
\end{titlepage}

\vfill
\clearpage

\begin{Large} \begin{center} H1 Collaboration \end{center} \end{Large}
\begin{flushleft}
%\input{h1auts}
%----------------------------------------------------------------------
%--Status: 26/6/95
%----------------------------------------------------------------------
 S.~Aid$^{14}$,                   %HAM2-PD      8/93        Aid
 V.~Andreev$^{26}$,               %LPI-PD                   Andreev
 B.~Andrieu$^{29}$,               %ECPL-PD                  Andrieu
 R.-D.~Appuhn$^{12}$,             %DESY-PD     4/92         Appuhn
% C.~Arnault$^{28}$,              %ORSA-TP                  Arnault
 M.~Arpagaus$^{37}$,              %ZUTH-LEFT   4/95         Arpagaus
 A.~Babaev$^{25}$,                %ITEP-PD                  Babaev
 J.~B\"ahr$^{36}$,                %ZEUT-PD                  Baehr
% E.~Banas$^{7}$,                 %CRAC-TP    6/93          Banas
 J.~B\'an$^{18}$,                 %KOSI-PD                  Ban1
 Y.~Ban$^{28}$,                   %ORSa-ST                  Ban2
 P.~Baranov$^{26}$
% ...

\end{flushleft}
\begin{flushleft} {\it
%     H1 Institutes as appearing on publications
%     H1 Institutes as appearing on publications
% ...
$\:^2$ III. Physikalisches Institut der RWTH, Aachen, Germany$^ a$ \\
% ...
$ ^{12}$ DESY, Hamburg, Germany$^a$ \\
% ...
$ ^{14}$ II. Institut f\"ur Experimentalphysik, Universit\"at Hamburg,
                                     Hamburg, Germany$^ a$  \\
% ...
$ ^{18}$ Institute of Experimental Physics, Slovak Academy of
                Sciences, Ko\v{s}ice, Slovak Republic$^ f$\\
% ...
$ ^{25}$ Institute for Theoretical and Experimental Physics,
                                                 Moscow, Russia \\
 $ ^{26}$ Lebedev Physical Institute, Moscow, Russia$^ f$ \\
% ...
$ ^{28}$ LAL, Universit\'{e} de Paris-Sud, IN2P3-CNRS,
                            Orsay, France\\
 $ ^{29}$ LPNHE, Ecole Polytechnique, IN2P3-CNRS,
                             Palaiseau, France \\
% ...
$ ^{36}$ DESY, Institut f\"ur Hochenergiephysik,
                              Zeuthen, Germany$^ a$\\
 $ ^{37}$ Institut f\"ur Teilchenphysik,
          ETH, Z\"urich, Switzerland$^ i$\\
% ...
$ ^a$ Supported by the Bundesministerium f\"ur
                                  Forschung und Technologie, FRG
 under contract numbers 6AC17P, 6AC47P, 6DO57I, 6HH17P, 6HH27I, 6HD17I,
 6HD27I, 6KI17P, 6MP17I, and 6WT87P \\
% ...
$ ^f$ Supported by the Deutsche Forschungsgemeinschaft\\
% ...
$ ^i$ Supported by the Swiss National Science Foundation\\
   } \end{flushleft}
\newpage

%>>>>>>>>>>>>>>>>>>>>>>>>>>>>>>>>>>>>>>>>>
%>>>>>>>>>>>>>>>>>>>>>>>>>>>>>>>>>>>>>>>>>>>>>>>>>>>>
% add author list
%>>>>>>>>>>>>>>>>>>>>>>>>>>>>>>>>>>>>>>>>>>>>>>>>>>>>
%\newpage
%==============================================================================
\section{Introduction}

     Photon-proton processes
are traditionally classified according to the virtuality ($Q^2$)
of the photon.
For quasi-real photoproduction interactions
 $Q^2$ is close to zero, and
correspondingly  the photon is nearly
on mass shell. Interactions which involve
a $Q^2$ larger than a few GeV$^2$ are usually
termed
deep-inelastic scattering processes (DIS).
This distinction results mainly from the
  different theoretical
descriptions adopted for these processes.
Photoproduction  has turned out to be very
similar to hadron-hadron collisions and is described in a VMD-like
(Vector Meson Dominance) picture~\cite{sakurai},
where the photon is assumed to fluctuate into a vector meson before interacting
with the proton. Additional diagrams such as the direct coupling of the
photon to quarks in the proton and a pointlike contribution where the
photon splits into a $q\overline{q}$ pair are needed to accommodate  the
large tail
 observed
 in the transverse momentum
 distribution of produced particles~\cite{H1:incl}, but the
 majority of photoproduction collisions show
 features of soft low-$p_T$ processes as seen in hadronic collisions. For DIS
on the other hand,
the observation of scaling of the structure function in
early experiments suggested that this interaction could, in the infinite
momentum frame of the proton, be interpreted as a hard scattering process
in which a
 point-like virtual photon ``probes'' the structure of the hadronic
target.
The subsequently measured scaling violations are well
described by perturbative QCD.

However,
recent DIS measurements  at the high energy  $ep$
collider HERA and high precision  measurements from  fixed target
leptoproduction experiments have  shown that
the data in the low Bjorken-$x$ region  reveal
 properties from  soft interactions as well. Here $x = Q^2/2P\cdot q$ where
$P$ and $q$ are the
four momenta of the incoming proton and photon, respectively.
Elastic and other diffractive hadronic final states have been
produced~\cite{h1rapgap,disdif}
 and shadowing
on nuclear targets has been observed~\cite{shadowing}, phenomena
which  are well known from
hadron-hadron and real photon collisions~\cite{gvmd}.
These observations  suggest that the different treatment of
  low and high $Q^2$ interactions is somewhat artificial, and it is
therefore worthwhile testing a prescription which provides a
 natural  transition between these two classes.

   A qualitative approach for a smooth transition of the high
energy $\gamma^*p$~\footnote{In this paper
the  generic symbol $\gamma^*$ is used to denote a
colliding photon irrespective of the virtuality.}
 interactions over a large range of $Q^2$ results
from viewing the photon-proton collision in the proton rest frame and
the hypothesis that the photon can develop
 as a hadronic object before
interacting  with the nuclear target.
The time in which a real photon can fluctuate into e.g. a $\rho$ meson is given
by the Heisenberg uncertainty principle and amounts to
$\tau \approx 2\nu/M^2_{\rho}$ where $\nu$ is the photon energy in the
proton rest frame and  $M_{\rho}$ the mass of the $\rho$ meson.
   The lifetime
of a high energy  real photon to fluctuate
 into a hadronic state  is much longer than the time of the
strong interaction itself, and therefore
  this  picture is
%This picture is different from the infinite proton momentum frame
 generally applied to
describe  photoproduction interactions.
The same argument can be used    for  DIS
interactions at low $x$. In the proton rest frame the time in which the
virtual photon can fluctuate and stay in a hadronic state, e.g. a
quark-antiquark pair,
 is given by
$\tau \approx 1/(xM_p)$ ~\cite{tioffe,brodsky}, where
$M_p$ is the mass of the nucleon target.
 Thus, for small $x$, the virtual photon
can convert into a quark-antiquark
 pair and cover a distance which is long  compared to the   interaction length.
 For an $x$ in the range of
$10^{-2}-10^{-4}$ this distance is in the range of 10 to 1000 fm,
much larger than
the typical radius of the target. In the HERA kinematic range such $x$ values
can be reached for  $Q^2 \sim 10 $ GeV$^2$.
Therefore, at low $x$, a virtual photon
can stay
 in a hadronic state for a long time and interact
 with the target strongly, leading to a final state
 similar to the one in a   hadron-hadron collision.
This picture of DIS in the frame where the incoming photon fluctuates
into a hadronic system before interacting with the proton has already been
advocated in various  papers~\cite{nikolaev,alligned,kwiecinski}.
Note that this
%a hadron-like interaction of virtual photons
is not in
contradiction with   the traditional treatment of DIS as a point-like process,
where a virtual photon ``probes" the hadronic structure of the nuclear target.
In fact  both pictures are taken to be complementary, as discussed
in~\cite{brodsky}.

The
HERA collider
provides a unique opportunity to study the final state
of both photoproduction and DIS interactions at high energy and small
Bjorken-$x$: 26.7 GeV electrons collide with 820 GeV protons, yielding a
centre of mass energy $\sqrt{s}$ of 296 GeV. This allows for a  study of
DIS interactions for $x$ values down to $\sim 10^{-4}$~\cite{h1f2},
and photoproduction
collisions at a centre of mass energy of $\sim 200$ GeV~\cite{H1:incl}.
The data recorded with the H1 experiment are used to
study the transverse energy behaviour in photoproduction and
DIS interactions, and the analogy with hadronic collisions is checked.
Within this analogy $\gsp$ events can be sub-divided
in rapidity-space into
three regions, which differ in the mechanism by which the hadronic final
state is produced~\cite{bj73}: the proton fragmentation region,
the photon fragmentation region and a hadron plateau
spanning the rapidity interval between the
two. The height  of the hadronic plateau depends logarithmically
on the centre of mass energy of the $\gsp$ collision.
The region of the hadronic plateau is expected to be independent of the
nature
of the two incoming ``hadrons",
as was verified in
hadron-hadron collisions~\cite{hadver}.
Particle production in  the proton fragmentation region is also expected
to be very similar compared to  hadron-proton collisions,
 but the limited experimental
acceptance in this region does not allow this to be verified. The ``hadronic"
nature of the photon is assumed to change
with increasing $Q^2$, thus
 the photon
fragmentation region is  expected to change with $Q^2$.
%This change is very sensitive to the $Q^2$ evolution of the hadronic
%structure of the photon on arrival at the target.
%In order to test this hadron-like analogy
%we compare event shape of high $Q^2$
%DIS events with real photon collisions, where the hadron-like
%photon interaction is already established.

Also studied is
 the fraction of diffractive events in
DIS and photoproduction.
In hadronic collisions
    approximate factorization of  the cross section for high mass
    diffractive dissociation has been observed, in accordance with
   Regge theory predictions~\cite{triple}:
  the ratio of the  hadron diffractive dissociation to the total
  cross section is found to be approximately independent of the type of
   dissociating
  hadron and is thus the
 same for pions, kaons, protons and real
photons~\cite{triple2}.
Applying this factorization property
 to real and virtual photons leads to  the expectation that the
ratio of the
 photon diffractive dissociation cross section to the total cross
section
is  independent of the $Q^2$ of the photon, in the high dissociative mass
region.

%==============================================================================
\section{Detector Description}

A detailed description of the H1 apparatus can be found elsewhere~\cite{h1}.
In the following  the components of the detector relevant for
this analysis are briefly described.

The hadronic energy flow and the scattered electrons in DIS processes
are measured with a
liquid argon~(LAr) calorimeter and a
backward electromagnetic lead-scintillator calorimeter (BEMC).
The LAr calorimeter~\cite{larc}  covers the polar angular range
$4^\circ < \theta <  153^\circ$ with full azimuthal coverage, where
 $\theta$ is defined with respect to the proton
beam direction ($+z$ axis).
It consists of an electromagnetic section with lead absorbers
and a hadronic section with steel absorbers.
Both sections are highly segmented in the transverse and
longitudinal direction
with about 44000 cells in total.
The total depth of both sections
varies between 4.5 and 8 interaction lengths.
Test beam measurements of the LAr~calorimeter modules show an
energy resolution of $\sigma_{E}/E\approx 0.12/\sqrt{E\;[\GeVx]} \oplus 0.01$
for electrons ~\cite{elcern} and
 $\sigma_{E}/E\approx 0.50/\sqrt{E\;[\GeVx]} \oplus 0.02$  for
charged pions~\cite{h1calo3}.
The uncertainty in the absolute energy scale for electrons is
 $3\%$.
The absolute scale of the hadronic energy is presently known to $5\%$, as
determined from studies of the
transverse momentum balance in DIS events.

The BEMC (depth of 22.5 radiation lengths or 1
interaction length) covers the backward region of the detector,
$151^\circ < \theta < 177^\circ$.
A major task of the BEMC is to trigger on and  measure scattered electrons
in DIS processes with $Q^2$ values ranging from 5 to 100 GeV$^{2}$.
The BEMC energy scale for electrons is known to an accuracy of
$1.7\%$~\cite{bemc}.
Its resolution is given by
$\sigma_{E}/E = 0.10/\sqrt{E\;[\GeVx]} \oplus 0.42/E[\GeVx] \oplus 0.03$
\cite{f2pap}.

The calorimeters are
surrounded by a superconducting solenoid providing a uniform
magnetic field of $1.15$ T parallel to the beam axis in the tracking region.
Charged particle
tracks are measured in a central drift chamber and the forward
tracking system,
covering the polar angular range $ 7^\circ < \theta < 165^\circ$.
   The central chamber is interleaved with an inner and an outer double
layer of multi-wire proportional chambers (MWPC),
which were used for the trigger to select
events with charged tracks pointing to the interaction region.
A backward proportional chamber (BPC) in front of the BEMC with an angular
acceptance of $155.5^\circ < \theta < 174.5^\circ$ serves to
support electron identification
and to   precisely measure electron direction.
Using information from the BPC, the BEMC and the reconstructed event vertex the
polar angle of the scattered electron is known to better than 2 mrad.

A small angle detector (electron tagger),
which is part of the luminosity system, is a TlCl/TlBr
crystal calorimeter with an energy resolution
$\sigma_{E}/E = 0.1/\sqrt{E\;[\GeVx]}$. It is located at $z$ = --33 m
and accepts electrons from photoproduction processes
with an energy fraction between 0.2 and 0.8 with
respect to the beam energy and scattering angles  $\theta'< 5$~mrad
 ($\theta' = \pi-\theta$).

%==============================================================================
\section{Event Selection and Correction for Detector Effects}

The data used in this analysis were collected in 1993
%in collisions of 26.7~GeV electrons with 820~GeV protons at HERA,
and correspond to
an integrated luminosity of about  $0.3$ pb$^{-1}$.
The kinematic variables $Q^2, x$ and $y$
of the $ep$ collision are determined from
 the scattered electron:
$ Q^2 = 4\,E_e \, E'_e\cos^2(\theta_{e}/2)$ and
$ y = 1-(E'_e/E_e)\cdot \sin^2(\theta_{e}/2)$,
where $E_e$ is the energy of the incident electron
and  $E_e'$ and $\theta_e$ are
the energy and the polar angle of the scattered electron respectively.
The variable $y$ represents the fraction of the energy of the electron
transferred to the proton, in the proton rest frame.
The scaling variable $x$
is then derived via $x=Q^2/(ys)$, and the total hadronic invariant mass
is given by
$W^2=sy-Q^2$.

\begin{table}[thb]
\centering
\begin{tabular}{|c|c|c|} \hline
 photoproduction & low $Q^2$ DIS  & high $Q^2$ DIS \\
    sample      &      sample     &      sample \\ \hline
$Q^2<10^{-2}$~GeV$^2$ & $5<Q^2<100$~GeV$^2$ & $Q^2>100$~GeV$^2$  \\ \hline
$\theta_e-\pi < 5$~mrad & $157^\circ<\theta_e < 173^\circ$ &
           $10^\circ < \theta_e < 148^\circ$
                      \\ \hline
$0.3 < y < 0.5$ & $E'_e > 12$~GeV & $0.05< y < 0.7$    \\
 & $y > 0.05$      &    \\  \hline
\end{tabular}
\caption[~]
{\small
  Accepted  kinematic regions for the three sub-samples used in this analysis.
}
 \label{seltable}
 \end{table}

The data are classified into three event sub-samples
depending on the $Q^2$ range. A different detector component of H1
is used to detect
the scattered electron in each of these ranges. In the studies below
these  subsamples
get further sub-divided into samples with different $y$ or $Q^2$ values.

\begin{itemize}

\item
The photoproduction sub-sample ($Q^2<10^{-2}$~GeV$^2$)
consists of events where the scattered electron is detected in the
small angle electron detector.
To
avoid regions of low acceptance and to
facilitate the data correction procedure,
the kinematic range was further restricted to $0.3<y<0.5$.
The photoproduction  events
are triggered by a coincidence of an energy deposit in the small
angle electron detector
and at least one track pointing to the vertex region.
The track condition is derived from the cylindrical MWPC and
requires a $p_T\;\gsim\;200$ MeV/c.
% The energy deposited in the electron tagger
%is considered to be due to the scattered electron.

\item
The low $Q^2$ DIS sub-sample ($5<Q^2<100$~GeV$^2$) consists of events
where  the scattered electron is detected in the BEMC. The events
are triggered by requiring
a cluster of more than 4~GeV in the BEMC.
The most energetic BEMC cluster is taken to be
the scattered electron.

\item
For the high $Q^2$ DIS sub-sample ($Q^2>100$~GeV$^2$)
the scattered electron is detected in the LAr calorimeter.
The events
are triggered by requiring
a cluster of more than 5~GeV in the electromagnetic part of the LAr
calorimeter,
and no associated hadronic energy.
The electromagnetic cluster in the LAr calorimeter with the highest transverse
energy is considered  to be  the scattered electron.
\end{itemize}

Further details on the scattered electron identification procedure
in the BEMC and LAr calorimeters and in the small angle electron detector
can be found in~\cite{h1f2,h1alphas,h1gamp},
respectively.

The selected kinematic regions  for each sub-sample, as given
in Table~\ref{seltable}, are
chosen to ensure a large acceptance,
high trigger efficiency and,
for the DIS sub-samples, a small photoproduction background (less than 3\%).
For all three sub-samples the $z$ position of the
event vertex reconstructed from charged
tracks was required to be within $\pm30$~cm
of the nominal interaction point.
A minimum $y$ cut, $y > 0.05$ was imposed for the DIS sub-samples.
Events suffering from QED radiation or from a
badly reconstructed electron
are strongly reduced by requiring that they also fulfill
this cut if $y$ is calculated from the measured hadrons.
%the requirement
%$W^2> 4400$~GeV$^2$ (corresponding to $y > 0.05$),
%with $W$ determined from the measured hadrons, in order to
%avoid large smearing effects and reduce radiative corrections.

The final event samples contain 82850 photoproduction events,
15324 low $Q^2$ DIS events with
$5 < Q^2 < 100$~GeV$^2$ ($10^{-4} < x < 10^{-2}$),
692 high $Q^2$ DIS events with
$Q^2 > 100$~GeV$^2$ ($10^{-3} < x < 10^{-1}$).

The data are corrected for detector effects using
samples of Monte Carlo generated events which were fully
 simulated  in the H1 detector.
The PHOJET~\cite{phojet} generator for photoproduction
and the CDM~\cite{ariadne}
 (Colour Dipole Model) generator for DIS processes were used.
The  program  PHOJET  generates $\gamma p$ interactions, treating the
photon as a hadron-like object.
The   model used
to simulate the hadronic final states is similar to that used
 in the Monte Carlo program
DTUJET~\cite{dtujet} which simulates particle production
  in $pp$ and $\bar{p}p$ collisions up to very high energies.
%
%  In the minimum-bias event generator PHOJET~\cite{phojet},
%  the multiparticle non-diffractive final
%  states are constructed from a parameterization of the photon-proton
%  scattering amplitude in eikonal approximation using the two-component
%  Dual Parton Model~\cite{dpm}.  The realization of the model is similar
%  to the MC generator DTUJET~\cite{dtujet} simulating particle production
%  in $pp$ and $\bar{p}p$ collisions up to very high energies.
%
  The CDM Monte Carlo program generates DIS events, and uses
 the colour dipole
  model for QCD radiation in the hadronic final state.
  Here the final state is
  assumed to be a  chain of independently radiating dipoles formed by
  emitted gluons~\cite{dipole}. Since all radiation is
  assumed to  originate from the dipole
  formed by the struck quark and the remnant, photon-gluon fusion events
  have to be added and are taken from the QCD matrix elements~\cite{lepto}.
  Version 4.03 of the ARIADNE program was used for the CDM
  studies in this paper.

To estimate systematic uncertainties of the correction procedure
for DIS events
another model for the hadronic final state was used as well:
the MEPS (Matrix Elements plus Parton Showers) model~\cite{lepto}.
  This model incorporates
  QCD matrix elements up to first order, with additional
  soft emissions generated by adding leading log parton showers.
  Divergences of the matrix element are avoided
  by imposing a lower limit
 on the  parton-parton invariant masses,
  which  was parametrized as a function of $W$ such that it is always
   2 GeV above the region in phase space where the
   order $\alpha_s$ contributions
  would exceed the total cross section. More details on this implementation are
given in~\cite{h1gljet}.
%  instead of the default cut-off with constant $\ycut=0.015$.

%==============================================================================

\section{Results}

For comparisons of event properties  at different $Q^2$ values,
the $\gamma^*p$ centre of mass system (CMS) is chosen as frame of
reference.
The orientation of the CMS is such that the
direction of the proton defines the positive $z'$ axis.
Transverse quantities are defined with respect to the proton direction
in this frame.

\begin{figure}[htb]  \centering
 \Large \boldmath
 \begin{picture}(170,110)(-8,0)
\put( 26,98){\small $\average{Q^2}\sim 0 $ GeV$^2$}
\put( 26,92){\small $\average{Q^2}\sim 11 $ GeV$^2$}
\put( 26,86){\small $\average{Q^2}\sim 38 $ GeV$^2$}
\put( 26,80){\small $\average{Q^2}\sim 520 $ GeV$^2$}
\put( 115,5){$\eta^* $}
\put( 5,+42){\begin{sideways}
  $ 1/N \cdot dE_T/d\eta^*~~[$GeV$]$ \end{sideways}}
\epsfig
{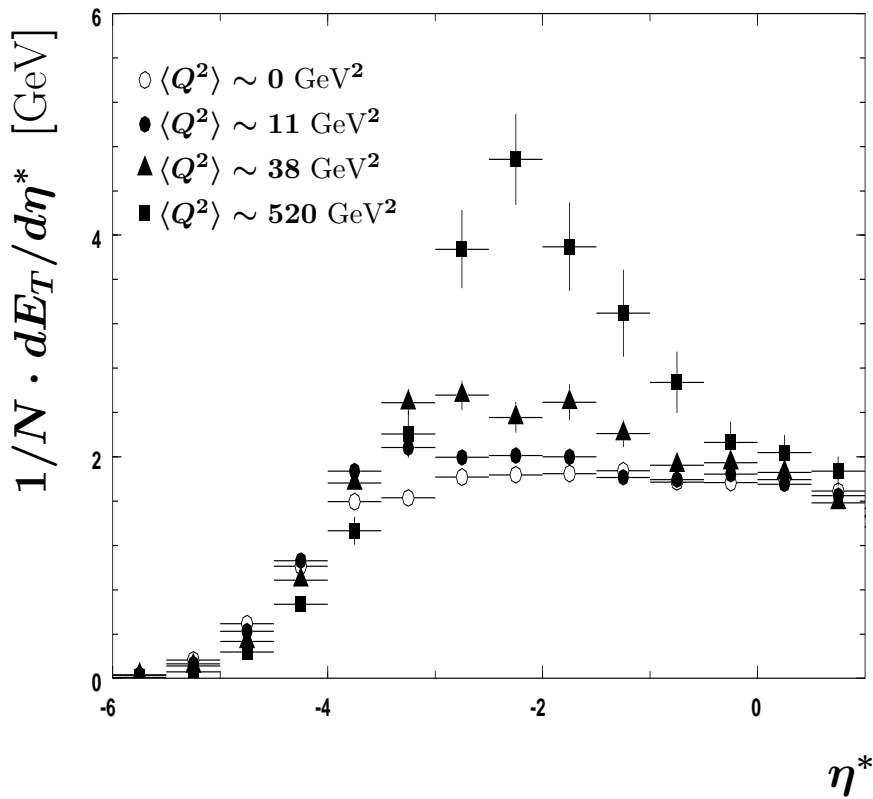}
\end{picture}
 \normalsize \unboldmath
\caption[~]
{\small
    The flow of transverse energy $E_T$ in the hadronic CMS as a function of
    pseudorapidity $\eta^*$ normalized to the number of events $N$.
    Photoproduction data (open circles) are compared with
    DIS data (full symbols: circles -- $\average{Q^2}\approx 11$~GeV$^2$,
    triangles -- $\average{Q^2}\approx 38$~GeV$^2$,
    squares -- $\average{Q^2}\approx 520$~GeV$^2$)
    in the same $y$ range $0.3 < y < 0.5$
    ($W\approx 185$ GeV).
}
\label{ETA}
 \end{figure}

The flow of transverse energy, $E_T$,
as a function of pseudorapidity $\eta^*=-\ln(\tan \frac{\theta^*}{2})$
in the CMS is shown in Fig.~\ref{ETA}.
Here $\theta^*$ is the polar angle of a particle in the CMS frame with
respect to the proton direction.
The transverse energy is calculated from the energy deposits in the
calorimeter cells.
For this study the  DIS data are restricted to the same kinematic region
$0.3<y<0.5$ as the photoproduction events.
 The $y$ range corresponds
to a  mean $\gsp$ collision energy of
$\sqrt{s_{\gamma p}} = W \approx 185$~GeV.
The corrections which have been applied to correct for detector
effects never exceed  30$\%$.

%\input{figures}
%..........................................................................
%
 \begin{figure}[htb]  \centering
 \Large \boldmath
 \begin{picture}(170,110)(7,0)
\put( 7,+45){\begin{sideways}
  $1/N\cdot dE_T/d\eta^* ~[$GeV$]$ \end{sideways}}
\put( 38,89){\small $-3<\eta^*<-2$}
\put( 38,96){\small $-0.5<\eta^*<0.5$}
\put( 33,82){\small -------- CDM}
\put( 33,76){\small -- -- -- MEPS}
\put(125,5){$Q^2~~[$GeV$^2]$ }
\put( 36,18){\large $\wr$}
\put( 37,18){\large $\wr$}
\put( 36,107){\large $\wr$}
\put( 37,107){\large $\wr$}
\put( 26,15){\small \bf 0}
\epsfig
{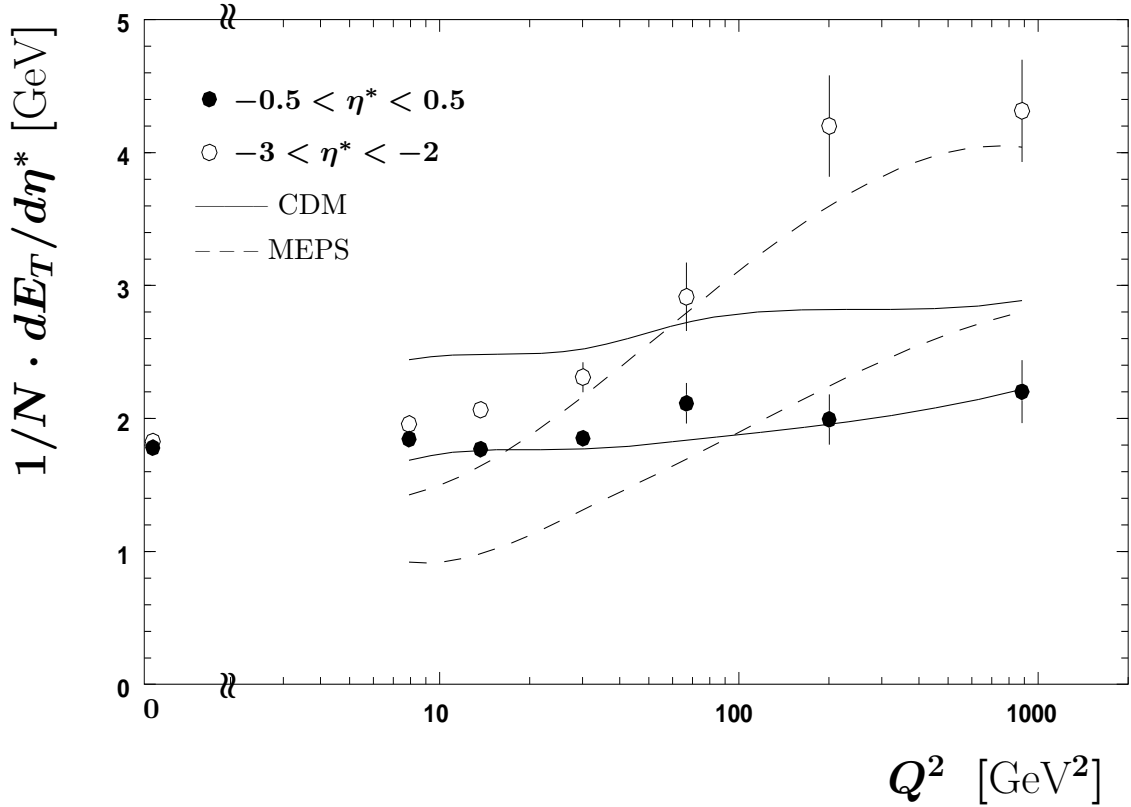}
\end{picture}
 \normalsize \unboldmath
\caption[~]
  {\small
The transverse energy per unit of pseudorapidity in
the CMS central region ($-0.5<\eta^*<0.5$, full circles) and in the photon
fragmentation region ($-3<\eta^*<-2$, open circles) as a function of $Q^2$
for $0.3 < y < 0.5$.
For comparison, the CDM (full line) and
MEPS (dashed line) models are shown. The lower curves correspond to
the CMS central region and the upper ones  to the photon
fragmentation region.
}
\label{Q2}
 \end{figure}
 \begin{figure}[htb]  \centering
 \Large \boldmath
 \begin{picture}(170,110)(7,0)
\put( 40,97){\large H1, $Q^2\approx~8~$GeV$^2$}
\put( 40,92){\large H1, $Q^2\approx~14~$GeV$^2$}
\put( 40,87){\large H1, $Q^2\approx~30~$GeV$^2$}
\put( 40,82){\large H1, photoproduction}
\put( 40,77){\large UA1}
\put( 40,72){\large AFS}
\put( 40,67){\large NA22}
\put( 7,20){\begin{sideways}
  $1/N\cdot dE_T/d\eta^*$ at $\eta^*=0~[$GeV$]$ \end{sideways}}
\put(125,5){$ W ~~[$GeV$]$ }
\epsfig
{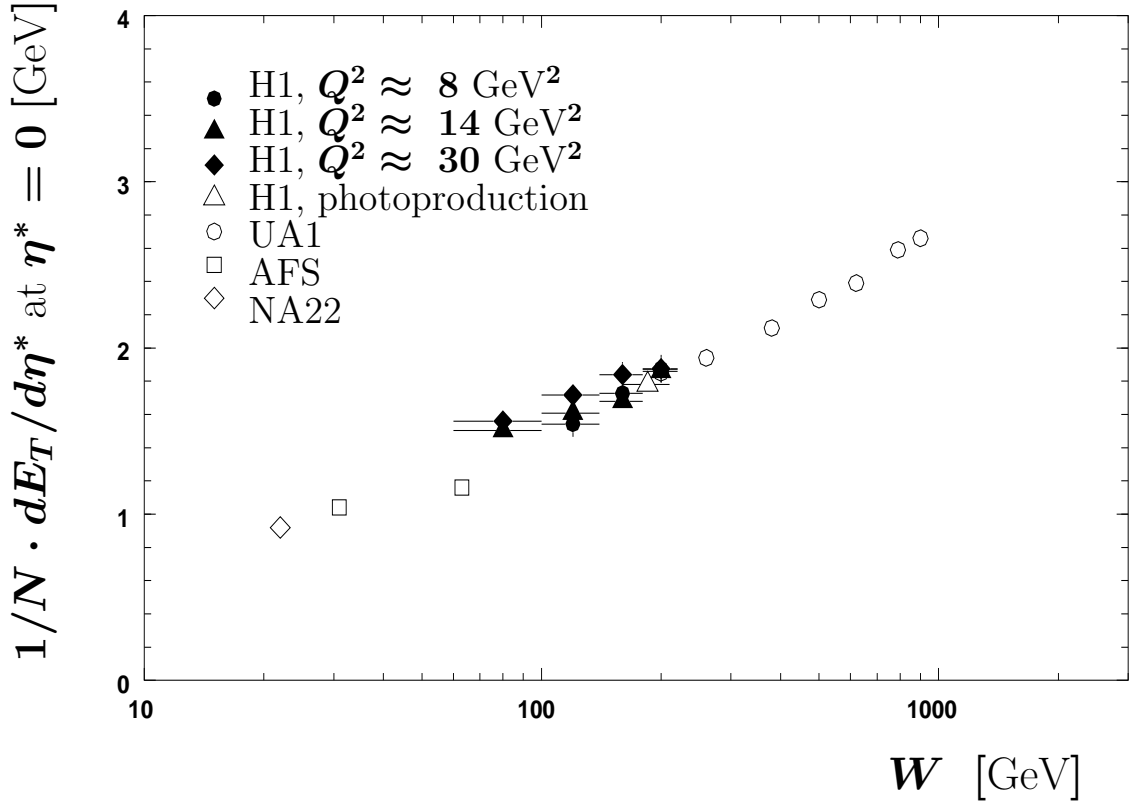}
\end{picture}
 \normalsize \unboldmath
\caption[~]
{\small
The transverse energy per unit of pseudorapidity
in the CMS central region
 as a function of the hadronic CMS energy.
The DIS data are compared with photoproduction data and with
 data from  hadron-hadron  collisions ($p\overline{p}$ for UA1;
$pp$ for NA22 and AFS).
Systematic point-to-point errors of $6\%$ and an overall scale
error of $9\%$ for the H1 data are not shown,
neither are the global scale uncertainties for the other experiments.
}
\label{W}
 \end{figure}
 \begin{figure}[htb]  \centering
 \Large \boldmath
 \begin{picture}(170,110)(7,0)
%
%>\put( 65,60){\Large $a)$}
%>\put(145,60){\Large $b)$}
\put( 25,100){\Large $a)$}
\put(105,100){\Large $b)$}
\put(  50,5){$E_T~~[$GeV$]$}
\put( 5,+30){\begin{sideways}
  $ 1/N \cdot dN/dE_T~~[$GeV$^{-1}]$ \end{sideways}}
\put( 140,5){$\eta^*_{max} $}
\put(85,+45){\begin{sideways}
  $ 1/N \cdot dN/d\eta^*_{max}$ \end{sideways}}
\epsfig
{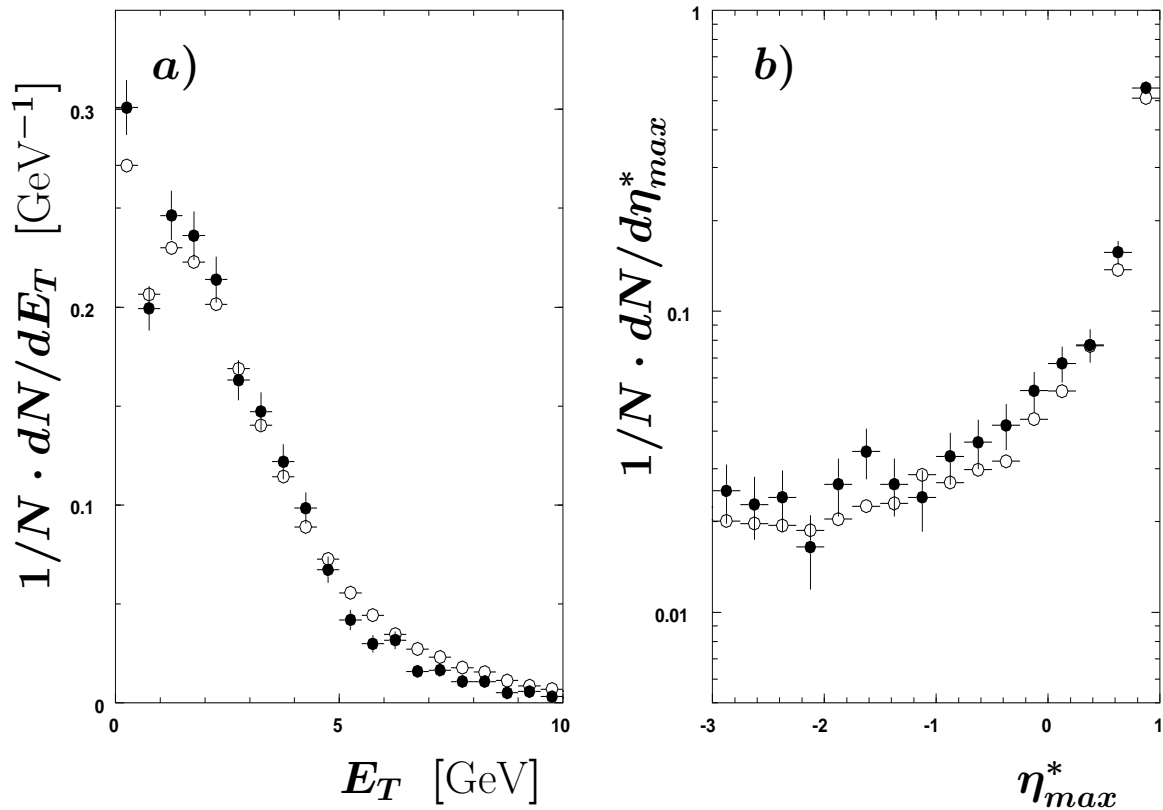}
\end{picture}
 \normalsize \unboldmath
\caption[~]
     {\small
$(a)$ distribution of the uncorrected transverse energy per unit
of pseudorapidity in the central region $(-0.5<\eta^*<0.5)~$,
$(b)$ $\eta^*_{max}$ distribution normalized to the total number of events.
Open circles -  photoproduction data, full circles - DIS data
with~$10<Q^2<100$~GeV$^2$ and $0.3<y<0.5$.}
\label{ET-MAX}
 \end{figure}
Only statistical errors are shown in Fig.~\ref{ETA}.
For the comparison of the $E_T$ spectra at
the four different $Q^2$ values it
is important to identify the relative systematic uncertainty, so
the systematic errors on each ``$Q^2$ point"
 at  a given  $\eta^*$ value are considered
in two parts.
Firstly there is a point-to-point systematic error, which has a different
effect on each of the four measurements,
 accounting
for  kinematically dependent systematic effects. Secondly there is
 an  overall scale error which affects all four points
at a given $\eta^*$ value in the same way.
The main source of the point-to-point systematic error
is the model dependence of the correction for detector effects.
This dependence was investigated with the
CDM and MEPS models and leads to a  point-to-point error of
$6\%$\footnote{In the central region the correction for photoproduction
events determined with PHOJET is close to the one for DIS events determined
with MEPS at $<Q^2>\sim 11$ GeV$^2$}.
%PHOJET) were compared and found to give the same correction factors for
%detector effects within 6\%)}.
Additionally, the model dependence contributes
$6\%$
to the overall scale error at each $\eta^*$ value.
Further contributions to the overall scale error
arise from the LAr and BEMC calorimeter calibration
($5\%$ and $20\%$ respectively), and
from details of the analysis method, such as the calorimeter noise
treatment,
the clustering scheme for the calorimeter cells,
and the accuracy of the simulation of the calorimeter
response,
affecting   the results by  $5\%$
in the region close to the proton direction  ($\eta^* > -1$).
All these contributions
give an overall scale error varying
from $8\%$ in the photon fragmentation region ($-3.5 < \eta^* < -1$) to
$9\%$ in the central region  ($\eta^* > -1$), and
up to $20\%$ in the region $\eta^* < -3.5$, which does not affect the
results of
the present analysis.
This discussion on the systematic errors is also valid
for data shown below.

Fig.\ref{ETA} demonstrates that
the transverse energy flow
exhibits a strong increase  in the photon fragmentation region
($\eta^* < -1$),
from $Q^2=0$ (photoproduction) to $Q^2 \approx 500$~GeV$^2$ (DIS),
 while in the central region the level of $E_T$ remains
almost the same.
This can be taken as evidence that the influence   of the $Q^2$
 of the photon on the $E_T$ flow diffuses
away quite quickly towards the central region.
Such behaviour is expected if the central region corresponds
to the hadronic plateau, discussed in the introduction.
In order to quantify the change of $E_T$ with $Q^2$,
 the value of
$E_T$ per unit of pseudorapidity ($E_T$ density)  for an $\eta^*$ slice
in the central
pseudorapidity region ($-0.5<\eta^*<0.5$) and
an $\eta^*$ slice in the photon fragmentation
region ($-3<\eta^*<-2$) is shown in Fig.~\ref{Q2}.
The data clearly  demonstrate
that the transverse energy in the central region is
essentially independent of the
 $Q^2$ of the photon, but increases significantly
in the photon fragmentation region.
However,  in the photon fragmentation region
the difference in the  $E_T$ density for DIS and
photoproduction
data is significant only for large $Q^2$ values, $Q^2 > 20 $ GeV$^2$.
The level of transverse energy
in the photon fragmentation region for interactions with $Q^2 < 10 $ GeV$^2$
is  quite similar, roughly independent of the transverse size,
$Q^2$, of the photon.
Since the former can be described by perturbative calculations, this
observation hints that the result of some
predicted features of the final state may be
transportable to the hitherto
assumed non-perturbative ``VDM region" at $Q^2 = 0$.

In Fig.~\ref{Q2}
 the results are compared with  predictions from the CDM and
MEPS models, as described in the previous section.
The $E_T$ density
 predicted by the CDM model is found to be independent of
$Q^2$, which is in accordance with the data in the central region, but
 does not
reproduce the data in the photon fragmentation region.
In this version of the CDM model the scale for QCD radiation is given by
the $p_T^2$  of the radiated gluons, which is limited by $W^{4/3}$.
Since  data  corresponding to
 the same $W^2$ region are selected,
  the predictions for the
$E_T$ density are similar for  the different $Q^2$ data samples.
A new version of the CDM model (version 4.06)\cite{leif2} partially
corrects for this defect in the photon fragmentation region, but still does
not yield a good description of the data.
 The MEPS model on the other hand describes the behaviour
of the transverse energy flow in the photon
fragmentation region but fails in the central region.
Hence, in this model, the influence of the
photon virtuality extends too far in  rapidity away from
 the photon-$q\overline{q}$ vertex.
The comparison of absolute values of the $E_T$ density
 between the data and
models should be viewed with some caution since  diffractive events are not
included in these models. These events, which amount to $\approx 6\%$
of all events,
give rise to  a
rapidity gap~\cite{h1rapgap},
i.e.  no energy deposition
in the central region.
%, are not included in the MEPS and CDM models.
The exclusion of diffractive events however does not change the shape of
the distribution of the  $E_T$ density versus $Q^2$.
Comparisons of the energy flows for models and data with  rapidity gap
events  removed can be found in~\cite{h1bfkl}.

Within the hadronic picture,
the observation of a rise of the  $E_T$ density
with $Q^2$ in the photon
fragmentation region implies that,  on
arrival at the target, the virtual photon has a hadronic structure
(parton configuration) which  depends on~$Q^2$.
With increasing  $Q^2$ the transverse momentum of the constituent
partons in this hadronic structure increases.
A formal explanation of this effect based on QCD
calculations is given in~\cite{khoze}.

The central region however shows a more universal behaviour, independent
of the partonic nature of the colliding particles.
It is therefore interesting to compare our measurement with
high energy hadron-hadron collisions. In
Fig.~\ref{W} the  transverse energy density,
$dE_T/d\eta^*$, at $\eta^* = 0$ is shown for DIS, photoproduction,
$pp$ and $\overline{p}p$ interactions as a function of $W$
(= $\sqrt{s}$ for hadron-hadron collisions).
The results for
DIS  include  only the data with $Q^2 < 50$~GeV$^2$, where  enough
statistics is available
to cover a large  $W$ (or equivalently
$y$)  range, as given in Table 1.
The $W$ dependence of $dE_T/d\eta^*$  observed in DIS processes at low $Q^2$
(and low $x$) agrees with
 the $W$ interpolation between  measurements
 from $p\bar{p}$~\cite{ua1mb} and
$pp$~\mbox{\cite{ua1mb,na22}}
collisions.
Also the photoproduction data show the same level of $E_T$ for the
$W$ value of this data sample.
This observation is consistent with  the ansatz of the  analogy of
$\gamma^*p$ interactions with  real photon-hadron and
hadron-hadron collisions.

In Fig.~\ref{ET-MAX}a
the distribution of (uncorrected)
 transverse energy summed over a
unit of pseudorapidity in the central region ($-0.5<\eta^*<0.5$)
is shown for  photoproduction and for  DIS data
($10<Q^2<100$~GeV$^2$, $0.3<y<0.5$).
The comparison shows that in the central region not
only the mean $E_T$,
but also the energy spectra themselves, are quite similar.
This apparently holds down to very low $E_T$ values,
where diffractive dissociation processes are expected to dominate
(see peak at $E_T=0$).
Since
for  $\eta^* = 0$  essentially the
same detector regions are used  in
  DIS
and photoproduction interactions
the agreement is not influenced by detector effects.
 Following this observation, it is interesting  to compare
the fraction of diffractive events in
 photoproduction and DIS.

An effective way to detect  diffractive processes at HERA is
 via  the $\eta^*_{max}$ variable, which
is the maximum pseudorapidity in the $\gsp$ CMS frame,
of a reconstructed track or  calorimetric
cluster with an energy larger than 400 MeV observed in the detector.
In diffractive events $\eta^*_{max}$ indicates the maximum pseudorapidity of
secondary hadrons from photon fragmentation and is related
to the fraction $x_p$
of the initial proton momentum carried by the diffractively scattered
proton via $\eta^*_{max} \sim \ln(1-x_p) + C$~\cite{kaidalov}.
The comparison of the $\eta^*_{max}$ distributions for DIS and photoproduction
interactions is shown in Fig.~\ref{ET-MAX}b.
The figure shows spectra which fall off rapidly
from $\eta^*_{max} = 1$ with
 decreasing $\eta^*_{max}$. This part of the spectra
 can be described by non-diffractive DIS and
photoproduction interactions as demonstrated in~\cite{h1rapgap,h1diffphoto}.
For $\eta^*_{max} \sim -1$ the
spectra level off to a constant  plateau. The events with this and
lower  $\eta^*_{max}$ values have been shown to originate predominantly from
a diffractive mechanism~\cite{h1rapgap2,h1diffphoto}. Below
 $\eta^*_{max}<-3 $ the photoproduction and DIS data are
 affected differently by the detector
acceptance and therefore not included  in this analysis.
As a consequence of this cut the low mass resonance region is avoided as well.
For diffractive events
the distribution $1/N\cdot dN/d\eta^*_{max} \sim
(1/\sigma_{tot})(d\sigma/d(1-x_p))(1-x_p)$  is expected to be roughly
independent of $\eta^*_{max}$ if the
diffractive cross section $d\sigma/d(1-x_p)$ is approximately
inversely proportional to $(1-x_p)$, see ref.~\cite{h1rapgap2} \footnote
{It was shown that for DIS events $d\sigma/d(1-x_p)$
is proportional to $(1-x_p)^n$ with $n=1.19\pm0.06\pm0.07$}.
This lack of dependence of $1/N\cdot dN/d\eta^*_{max}$ on $\eta^*_{max}$
is  itself often taken as the signature of diffraction~\cite{bjdiff}.

   Fig.~\ref{ET-MAX}b shows clearly that indeed at low $\eta^*_{max}$ both for
DIS and photoproduction the differential distribution
$1/N\cdot dN/d\eta^*_{max}$ is roughly independent of $\eta^*_{max}$.
Furthermore the
   relative contribution of the photon high mass diffractive dissociation
 for photoproduction and DIS interactions
    is found to be the same to within 15-20\%.
This agreement is not affected by detector effects since
for a given $\eta^*_{max}$ value largely the
same detector regions are explored  in
  DIS
and photoproduction interactions.
This observation is at the same level of agreement as measurements made for
hadron-hadron interactions, and is
 in accord  with the expectation of
    approximate factorization of high mass
    diffractive cross sections, as  observed in hadron collisions and
  explained with triple Regge phenomenology~\cite{triple}:
  the ratio of the differential diffractive cross section,
 $d\sigma/d(1-x_p)$, to the total
  cross section is approximately independent of the type of
  dissociating
  hadron.
%  It is easy to show$^{12}$ that the flat part of $\eta^*_{max}$ spectrum
%  presented in Fig.~\ref{ET-MAX}b)
%  corresponds to this ratio. Therefore, a close similarity of
%  $\eta_{max}$ spectra for DIS and photoproduction suggests that  and is found
%%to be
%the same for pions, kaons protons and real photons.
%  the factorization rule holds true also for virtual photons.
The  fact that   this factorization rule also seems
to hold for   virtual photon
  interactions gives an additional
  argument in favour of the validity of a universal
  hadron-like description of low-$x$ DIS, real photoproduction, and
  hadron-hadron collisions.

%==============================================================================
\section{Conclusions}

   The comparison of  photoproduction and low-$x$ DIS data
  at the $ep$ collider HERA  reveals  striking similarities
in the energy flow of the hadronic final state
and  relative rate of high mass diffractive
dissociation of the photon.
The $W$ dependence of the
 transverse energy density in the central rapidity region is found to
be similar to that seen in high energy hadron-hadron collisions.
These findings are consistent with the
hadronic  picture of the photon which can therefore be considered to be
complementary to the conventional deep-inelastic scattering picture, even
 at large $Q^2$ values.
This picture gives a description of
 the transition region from the high $Q^2$
perturbative region to the low $Q^2$ non-perturbative region, and provides
a basis for a universal description of hadron-hadron, real photon-hadron
and virtual photon-hadron high energy interactions.

%==============================================================================
\section*{Acknowledgements}
We are grateful to the HERA machine group whose outstanding efforts
made this experiment possible. We appreciate the immense
effort of the
engineers and technicians who constructed and maintained the detector.
We thank the funding agencies for their financial support of the
experiment. We wish to thank the DESY directorate for the support
and hospitality extended to the non-DESY members of the collaboration.
We thank further  J. Bartels, A. Kaidalov, J. Kwiecinski and
M. Ryskin for useful
discussions.

%==============================================================================
{\Large\normalsize}
\begin{footnotesize}

\end{footnotesize}

\end{document}